\documentclass[prb,preprint,endfloats]{revtex4}

\def\nle{\ \raise.3ex\hbox{$<$}\kern-0.8em\lower.7ex\hbox{$\sim$}\ }
\def\nge{\ \raise.3ex\hbox{$>$}\kern-0.8em\lower.7ex\hbox{$\sim$}\ }

\usepackage{bm}
\usepackage{graphics,epsf}
\usepackage{array}
\usepackage{amsmath}
\usepackage{amssymb}

\begin{document} 
\title{Brownian Dynamics Studies on DNA Gel Electrophoresis. II.\\
`Defect' Dynamics in the Elongation-Contraction Motion}
\author{Ryuzo Azuma}
\affiliation{Institute for Solid State Physics, The university of Tokyo
5-1-5 Kashiwanoha, Kashiwa, Chiba 277-8581, Japan}
\altaffiliation[Present address: ]{Genomic Sciences Center, RIKEN,
1-7-22 Suehiro, Tsurumi, Yokohama, Kanagawa, 230-0045, Japan}
\date{\today}

\begin{abstract}
By means of the Brownian dynamics (BD) method of simulations we have
 developed, we study dynamics of individual DNA undergoing 
constant field gel electrophoresis (CFGE), focusing on the relevance of
 the `defect' concept  due to de~Gennes in CFGE. The corresponding
 embodiment, which we call {\it slack beads} (s-beads) is explicitly
 introduced in our BD model. 
In equilibrium under a vanishing field the distance 
between s-beads and their hopping range are found to be 
randomly distributed following a Poisson distribution. In the strong 
field range, where a chain undergoes the elongation-contraction motion, 
s-beads are observed to be alternatively annihilated in elongation and 
created in contraction of the chain. On the other hand, the 
distribution of hopping range of s-beads does not differ much from that 
in equilibrium. The results indicate that the motion of the chain 
elongated consists of a huge number of random movements of s-beads. 
We have also confirmed that these features of s-beads agree 
qualitatively with those of s-monomers in the extended bond fluctuation 
model (EBFM) which we recently proposed. The coincidence of the two 
simulations strongly supports the stochastic semi-local movement of 
s-monomers which we {\it a priori} introduced into the EBFM.
\end{abstract}
\maketitle

\section{Introduction}

Gel electrophoresis is a major technique which separates
polyelectrolytes according to their size. Among them gel electrophoresis
of DNA is one of the most interesting problems of statistical physics of
polymers. Computer simulations~\cite{deut88,deut89} and 
experiments~\cite{masu93,oana94} reported a peculiar behavior of
individual DNA that they exhibit contracted and extended forms
alternatively under a constant field. Recently we have developed a
Brownian dynamics (BD) method which simulates electrophoresis of DNA in
a 3 dimensional (3D) space by a chain of electrolyte beads. Its details
have been reported in our preceding paper (hereafter refereed to as I),
where we have demonstrated `quasi-periodic' time evolution of the
velocity of the center of mass, $v_{\rm G}(t)$, and of the radius of the
longer principal axis, $R_l(t)$, of the chain.  The present paper is 
devoted to clarify the role of `defects' introduced by de 
Gennes~\cite{deGe71} on the elongation-contraction motion. 

The concept of `defect' proposed by de~Gennes~\cite{deGe71} is useful 
to investigate dynamics of individual polymers confined to a space with
entanglement such as gel electrophoresis of DNA. A defect represents a 
relatively loose portion, or a slack segment, of a polymer and is
assumed to behave as a particle of ideal gas. He showed that its
diffusion in the tube made of other polymers leads to the 
famous growth law $\phi(t)\propto t^{1/4}$ in the short time regime, 
where $\phi(t)$ is the mean square displacement of a segment in 
equilibrium. 

Electrophoresis of a chain in an ideal regular network of other fixed
chains has been extensively studied by Monte Carlo (MC) 
simulations.~\cite{cruz86,duke90,krem89} Among them, the repton model
proposed by Rubinstein is important.~\cite{rubi87} It simulates dynamics
of a chain only by stochastic movements of reptons each of which can be
regarded as embodiment of a defect. The model is applied to
simulation of the constant field gel electrophoresis (CFGE) by 
Duke.~\cite{duke90} It was shown that the mobility is inversely
proportional to the number of reptons in a very weak field region. The
relation was consistent with what the reptation theory predicted under
the vanishing field limit.~\cite{lerm82,lump82} Under finite fields, 
however, the results derived from the repton model qualitatively 
differ from  those derived from the biased reptation 
theory~\cite{lump85}: the mobility plateau is reached at lower fields 
in the former model than those predicted by the latter theory. Duke also 
observed snapshots of tubes which look like hooks, and argued that once 
a chain is hooked around a gel fiber, both its arms tend to stretch and 
rarely retract. This is because defects prefer to move in direction of 
the field and to create new tube segments which in turn prefer to align 
to the field. The observed fact that the chain rarely exhibits a 
retraction is considered to be an artifact of the MC method which uses 
local hopping rules. The incorporation of a hopping rule which allows 
non-local movement of defects was done by Duke and Viovy in their 
advanced version of the repton model.~\cite{duke92,duke92b} Simulations 
of chains with over a thousand segments revealed that a chain exhibits 
ramified configurations where `hernias' are created and annihilated under 
a constant field.~\cite{duke92} 

Recently, we have investigated gel electrophoresis and diffusive dynamics 
of a chain in a space with immobile obstacles by making use of an extended 
version of the bond fluctuation model (BFM).~\cite{azuma99a} The 
conventional BFM (CBFM)~\cite{carm88} is known as an efficient model
to simulate general problems involved in polymers. However, it is faced 
with a problem which is similar to that of the repton model mentioned 
above: the chain motion extremely slows down when it is applied to 
CFGE under relatively large fields.~\cite{krem89} To overcome this
problem, we have empirically incorporated semi-local dynamics of slack
parts of a chain. The latter correspond to `defects' due to de Gennes
and were named s-monomers. They are explicitly defined in each
configuration appearing in the CBFM. We have called such a BFM with 
introduction of semi-local movements of s-monomers the extended BFM
(EBFM). Both the DV model and the EBFM incorporate nonlocal movements of
slack portions of DNA. There are, however, the following difference
between the two methods.  In the DV model any defect is not created or
annihilated except at the extremity, whereas in the EBFM, s-monomers are
continually created and annihilated anywhere in the chain, since it
involves the CBFM trials of conventional local movements of all monomers.

Using the EBFM it has been confirmed that the trapping problem
of the CBFM under large fields is overcome and the field-dependence of 
mobility is reproduced qualitatively as observed by the
 experiment.~\cite{herb87} Furthermore, quasi-periodic behavior, or
a damped oscillation, is observed in the correlation function of $R_l(t)$
 in 3D~\cite{azuma00} as it has been observed in the 
experiment~\cite{masu93,oana94} as well as by the BD calculations in 2D
(without the excluded volume effect)~\cite{matsu94} and in 3D (with the 
effect).~\cite{ourI}

The defect is thus a quite important concept to understand DNA dynamics
in gel. The purpose of the present paper is to clarify the microscopic
bases of the defect dynamics, thereby presenting a microscopic support 
of the EBFM. For this purpose we examine the corresponding object in a
chain, i.e., slack beads (abbreviated as s-beads) by means of the BD 
method. The work which is similar to this approach has been done by 
Masubuchi et~al.~\cite{masu97} in their 2D BD study on the biased
sinusoidal field gel electrophoresis (BSFGE). Looking at time evolution
of snapshots of kinks which correspond to slack segments, they have
found that a portion with less kinks develops near the segment which is
hindered by an obstacle, and that the portion gradually shifts as the
chain slides off the obstacle. They have argued that the growth of kinks
plays a key role on the decrease of the mobility observed in chains with
particular sizes under BSFGE. 

In the present paper we examine static and dynamic properties of defects
much more in details by means of our BD method for a chain in a 3D
space. As described in our accompanying paper, hereafter referred to as
I,~\cite{ourI} the method 
incorporates semi-microscopic ingredients of dynamics of a
polymer, i.e., the excluded volume effect, the non-linear elastic force 
which keeps the distance between the two adjacent beads (segments) and
the random force from a solvent. In each configuration of the chain
simulated by the BD method s-beads, or `defects', are explicitly
specified  in a quite similar way as the s-monomers have been defined in
the EBFM. Then we investigate various statistics associated with the
s-beads, or more explicitly, portions between the neighboring
s-beads which we call extended parts of the chain. 

In particular, we
have concentrated on the normalized histogram of the extended parts of
length $n$, $r_n(t)$, and that of their displacement along the chain
direction, $r_{\Delta u}(t)$. In equilibrium under a vanishing field it
is found that both $r_n(t)$ and $r_{\Delta u}(t)$ obey Poisson
distributions as expected. In the elongation-contraction motion under 
CFGE, $r_n(t)$ exhibits a peculiar time evolution reflecting the
elongation and contraction of the chain. The histogram $r_{\Delta u}(t)$,
on the other hand, does not fluctuate correspondingly to $r_n(t)$ but
exhibits random hopping nature rather similar to the one observed in
equilibrium. These results indicate that defects move quite
stochastically even in the {\it deterministic} regime where the chain
elongation-contraction motion looks apparently deterministic. 
These features of $r_n(t)$ and $r_{\Delta u}(t)$ are confirmed to be
in qualitative agreement with those observed for s-monomers in the EBFM.

The organization of the present paper is as follows. In the next section
we briefly explain our numerical method, and discuss equilibrium dynamics
of s-beads in Sec.~\ref{sec:tb-eq}. Characteristic features of $r_n(t)$
and $r_{\Delta u}(t)$ in the field range where a chain exhibits
elongation-contraction motions are examined in Sec.~\ref{sec:def-dyn}. 
The corresponding results obtained by the EBFM are discussed in
Sec.~\ref{sec:ebfm}, and the final section is devoted to the conclusion. 

\section{Numerical Method}\label{sec:method}

The BD method we adopt in the present work is to solve the following 
Langevin equation of motion for a chain of spherical beads in a 
continuous 3D space,
\begin{equation}
\zeta \dot{\bm x}_i = q{\bm E}_{\rm b} 
+ \sum_\alpha \sum_j {\bm S}^\alpha_{ij} 
+ \sum_\alpha \left[ {\bm F}^\alpha_i-{\bm F}^\alpha_{i-1}\right] 
+ {\bm f}_i,
\label{eqn:lange}
\end{equation}
where ${\bm x}_i$ is the position of the $i$-th bead, and $\zeta$, $q$, 
and ${\bm E}_{\rm b}$ are the viscosity, the charge of a bead and 
the (bare) external field, respectively. 
The second and the third terms in r.h.s. are the constraining forces
required for the beads to satisfy the conditions,
\begin{eqnarray}
\|{\bm x}_{ij}\|=&\|{\bm x}_{j}-{\bm x}_i\|&>1\label{eqn:cexcl}\\
\|{\bm l}_i\|=&\|{\bm x}_{i+1}-{\bm x}_i\|&<\sqrt{2}.\label{eqn:cbond}
\end{eqnarray}
putting the bead diameter unity. The last term in eqn.~\ref{eqn:lange} 
stands for the random force from a solvent. Explicitly we adopt the 
following form for it
\begin{equation}
{\bm f}_i(t) = \sum_n {\bm f}_i^n \delta (t-n\Delta t),
\label{eqn:rndm}
\end{equation}
where $\Delta t$ is regarded as the average period of random forces. 
The distribution of ${\bm f}_i^n$ is characterized by the averages
\begin{eqnarray}
\langle {\bm f}_i^n \rangle &=& {\bm 0}\\
\langle {\bm f}_i^m{\bm f}_j^n\rangle &=& 
2\zeta k_{\rm B}T\Delta t\delta_{ij}\delta_{mn}.\label{eqn:chi}
\end{eqnarray}
Here $k_B$ and $T$ are the Boltzmann constant and the absolute 
temperature, respectively. 

The details of solution of the above set of equations have been explained 
in our accompanying paper I. As a representative of `defect', or a loose
portion of a chain in the BD calculation we define a slack bead (s-bead). 
For any sequences of three beads, if the distance 
between the first and the third is not greater than $\sqrt{2}$, the 
second one is regarded as an s-bead. This definition of `defect' is the 
same as that of s-monomers in a 3D version of the EBFM we investigated 
before~\cite{azuma00} (see in Sec.~\ref{sec:ebfm}). We call a sequence of 
beads between neighboring s-beads an extended part of the chain. 

\section{`Defects' in equilibrium}\label{sec:tb-eq}

In equilibrium under a vanishing field, `defects' are considered to be 
distributed randomly along the polymer coordinate. Correspondingly, in 
our BD analysis, s-beads are randomly distributed throughout the chain. 
Then the `extended' parts consisting of $n$ beads is expected to obey 
a distribution
\begin{equation}
\Psi(n)=\frac{(M-n)(1-\rho)^n}{\sum^M_{i=1}(M-i)(1-\rho)^i},
\label{eqn:rpsin}
\end{equation}
where $M$ is the total number of beads of the chain and $\rho$ the density 
of s-beads. For $M\gg n \gg 1$, $\Psi(n)$ reduces to a Poisson distribution:
\begin{equation}
\Psi(n)\cong \rho(1-\rho)^n=\rho e^{-n\log(\frac{1}{1-\rho})}.\label{eqn:psin}
\end{equation}

Next let us consider $\Delta u$, displacement of the tangential component,
of extended parts between neighboring s-beads. In our BD calculation,
it is evaluated as
\begin{equation}
 \Delta u\equiv \left|\sum_{i=i_m}^{i_{m+1}}\Delta s_i\right|, 
\quad{\rm with}\quad 
 \Delta s_i=\Delta{\bm x}_i\cdot\frac{{\bm x}_{i-1}-
{\bm x}_{i+1}}{\|{\bm x}_{i-1}-{\bm x}_{i+1}\|}
\label{eq:def-Du}
\end{equation}
where $\Delta{\bm x}_i$ is the displacement of the $i$-th bead in one MD 
step solving the Langevin equation~\ref{eqn:lange} and index $i_m$ 
denotes a figure which is numbered on the s-beads. From 
eqn.~\ref{eqn:lange} we may assume that the distribution of 
$\Delta{\bm  x}_i$ is essentially given by that of the random force, 
${\bm f}_i^n$, defined by eqn.~\ref{eqn:chi}. Then the distribution of 
$\Delta s_i$ is written as
\begin{equation}
\Xi(\Delta s_i)
=\sqrt{\frac{\zeta}{2\pi k_{\rm B}T}}
\exp(- \zeta\Delta s^2_i/2 k_{\rm B}T). 
\end{equation}
Using this $\Xi(\Delta s_i)$ and $\Psi(n)$ of eqn.~\ref{eqn:psin}, we can 
evaluate the distribution of $\Delta u$ as
\begin{eqnarray}
\Phi(\Delta u)&=&\int_0^\infty dn\Psi(n)
\prod_{i=1}^n\left[\int_{-\infty}^{\infty}\Xi(\Delta s_i) d\Delta s_i\right]\delta(\sum_{i=1}^n\Delta s_i-\Delta u)\nonumber\\
&=&\rho\sqrt{\frac{2\pi^2\zeta}{ k_{\rm B}T\log\frac{1}{1-\rho}}}
\exp \left[-\sqrt{\frac{\log\left(\frac{1}{1-\rho}\right)}
{k_{\rm B}T/2\zeta}}\Delta u\right].
\label{eqn:psidmeq}
\end{eqnarray}
Thus $\Phi(\Delta u)$ is expected to obey also a Poisson distribution
with the width $\Omega_{\Delta u}=
1/\sqrt{{\log(\frac{1}{1-\rho})}/({k_{\rm B}T/2\zeta)}}$.
The latter is proportional to $\sqrt{k_{\rm B}T}$ if the $T$-dependence
of $\rho$ is neglected.

In Fig.~\ref{fig:histdem}, we show the normalized histograms $r_n$ and 
$r_{\Delta u}$ in equilibrium for various $T$ which are 
obtained by the BD calculation. Here $r_n=N_n/\sum_n N_n$ where $N_n$
is the number that the extended parts of a length $n$ appear in the chain
configurations simulated, and $r_{\Delta u}$ is similarly defined.
In agreement with $\Psi(n)$ of eqn.~\ref{eqn:psin} $r_n$ obeys a Poisson
distribution. The exponential tail of $r_n$ at $k_{\rm B}T=1$ is
fitted to $e^{-n/\omega}$ with $\omega \equiv \langle n\rangle\cong 5.7$,
which yields
$\rho\cong 0.17$ through the relation $\omega=1/\log\left(\frac{1}{1-\rho}
\right)$. In the simulation we have also directly counted the number of 
s-beads, which yields the same s-bead density $\rho\cong 0.17$ as 
expected. The distribution $r_{\Delta u}$ of the tangential movement of 
the extended parts, which we regard as that of the s-beads, also obeys a 
Poisson distribution as seen in the same figure. Its exponential tail, 
specified by $\Omega_{\Delta u}$ as $e^{-n/\Omega_{\Delta u}}$, 
significantly depends on $T$. As shown in 
Fig.~\ref{fig:histdem-fit}, its dependence on $T$ is consistent 
with eqn.~\ref{eqn:psidmeq}. Even the pre-factor $0.30$ of the 
proportionality to $\sqrt{k_{\rm B}T}$ agrees with 
the predicted value $1/\sqrt{2\log{\frac{1}{1-\rho}}}\cong 0.37$ within
 $20\%$.

The results obtained above confirm the role of the `defects', or the 
s-beads, on the static and dynamic properties of a polymer in 
equilibrium. In particular, it is natural to regard 
$c\langle n\rangle \langle l\rangle$ as the `persistence length' of the 
chain, where $\langle n\rangle$ is the mean value of $n$ and 
$c$ is a numerical constant nearly equal to unity. In this context, we
note $\sqrt{2} > \langle l \rangle > 1$, where $\langle l \rangle$ is
the mean distance between two adjacent beads. 
In fact we have already argued in I 
that this interpretation yields an appropriate conversion factor between 
the length scales of the BD simulation and the experiment. Here we have
to emphasize that the value of $\langle n\rangle$ is almost independent
of $T$ in the range $k_{\rm B}T \nle 8$, as well as of $M$ and 
$a_{\rm gel}$ we have
examined. It is observed, however, to reduce about by a half when the
excluded volume effect between beads is discarded. This enlargement of
the `persistence length' is one of the important consequences of 
the excluded volume effect on the present BD chain in a 3D space.

\begin{figure}
\caption[]{
The normalized histograms $r_n$ and $r_{\Delta u}$ of a chain with
 $M=160$ under $E=0$ for various $k_{\rm B}T$ obtained by the BD
 calculation.  
}
\label{fig:histdem}
\end{figure}

\begin{figure}
\caption[]{\makebox[\textwidth][l]{
$\Omega_{\Delta u}$ versus $k_{\rm B}T$.}
}
\label{fig:histdem-fit}
\end{figure}

\section{`Defects' Dynamics under CFGE}\label{sec:def-dyn}

Under relatively strong fields DNA undergo the 
quasi-periodic behavior~\cite{oana94}: real time images of individual 
DNA exhibit elongated and contracted shapes alternatively. Corresponding 
behavior was observed in the previous 2D BD 
models~\cite{deut88,matsu94,mas95p} likewise by our present BD simulation 
which has been discussed in details in I. We show in 
Fig.~\ref{fig:WRL-BDC}, a normalized curve of $R_l(t)$, i.e.,
$\bar{R}_l(t-t_{\rm max})/\bar{R}_l(t_{\rm max})$ of chains with $M=160$ 
in a lattice of gel with the lattice distance of $a_{\rm gel}=20$ under 
a field $E \equiv qE_{\rm b}=0.032$. In order to get further insights 
into the mean elongation-contraction motion through the defect dynamics,
we have examined histograms of length of the extended parts $N_n(t)$ 
and of their tangential movement $N_{\Delta u}(t)$ at four representative 
time intervals, i.e., $\left[ t_{\rm min}-0.1D, t_{\rm min}\right]$,
$\left[ t_{\rm min}, t_{\rm min}+0.1D\right]$,
$\left[t_{\rm max}-0.1D, t_{\rm max}\right]$,
and $\left[t_{\rm max}, t_{\rm max}+0.1D\right]$,
as indicated in Fig.~\ref{fig:WRL-BDC}. Here 
$t_{\rm min}\equiv \overline{t_{\rm min:l}^n-t_{\rm max}^n}$
and $D\equiv \overline{t_{\rm min:r}^n-t_{\rm min:l}^n}$, where 
$t_{\rm max}^n$ is the time of local maximum in the $n$-th peak in 
$R_l(t)$, $t_{\rm min:r}^n$ and $t_{\rm min:l}^n$ are respectively 
local minima just before and after the local maximum at $t_{\rm max}^n$, 
and the overline stands for the average over peaks in $R_l(t)$. 
As for the procedure how to pick up the peaks in $R_l(t)$, the reader may 
refer to I. We denote the four intervals introduced here as Interval 1, 
2, 3 and 4, respectively. It is expected that Interval 1 
corresponds to the time range where the chain is in a coiled shape,
Interval 2 to that of a contracted shape when a U-shape conformation 
has just started to grow due to trapping by gel, Interval 3 to that of 
an elongated U-shape when the chain is going to get rid of the trap, 
and Interval 4 to that of an I-shape when it is going to retrieve another 
coiled shape. We cut out sequential $2000$ MD steps (mds) in each of the four 
intervals of each peak in $R_l(t)$ and accumulate data associated with 
them over 80 peaks whose $D$ are about $1.5\times 10^5$mds or larger.

\begin{figure}
\caption[]{
The mean peak shape
$\bar{R}_l(t-t_{\rm max})/\bar{R}_l(t_{\rm max})$ of $M=160$ with
$a_{\rm gel}=20$ under $E=0.032$.
In each of the four intervals indicated by the arrow,
histograms of $N_n(t)$ and $N_{\Delta u}(t)$ 
of a sequential $2000$ steps are accumulated. 
}
\label{fig:WRL-BDC}
\end{figure}

\begin{figure}
\caption[]{
Normalized histograms $r_n(t)$
represents characteristic instantaneous distributions of $n$ for the
intervals defined in Fig.~\ref{fig:WRL-BDC}. 
}
\label{fig:h-a-BDC}
\end{figure}

\begin{figure}
\caption[]{\makebox[\textwidth][l]{
Normalized histograms $r_{\Delta u}(t)$ in the four intervals.}
}
\label{fig:h-as-BDC}
\end{figure}

The normalized histograms $r_n(t)$ $(=N_n(t)/\sum_m N_m(t))$ and
$r_{\Delta u}(t)$ thus obtained are shown in Figs.~\ref{fig:h-a-BDC}
and \ref{fig:h-as-BDC}, respectively. They exhibit peculiar 
time evolution within the `elongation-contraction' period. One may at 
once notice from $r_n(t)$ that the chain is pulled by the field in 
Intervals 2 and 3. Particularly in Interval 3, $r_n(t)$ at 
$40 \nle n \nle 120$ becomes about $10^{-3}$ whose integrated weight 
reaches about $10\%$, i.e., among 10 extended parts one of them is as 
large as several tens. The latter is quite elongated and is considered 
to constitute the longer arm of the U-shaped chain. Interestingly, 
however, we can see from Fig.~\ref{fig:h-as-BDC} that $r_{\Delta u}(t)$ 
at $\Delta u$ larger than ten is always negligibly small. This means 
that sliding dynamics of the chain even in Interval 3 consists of a 
large number of defect movements whose $\Delta u$ are less than several 
times of the mean bead distance. Thus the dynamics even in the elongated
portion of the chain is considered to be quite stochastic when it is
looked at through $\Delta u$, i.e., changes in a unit of time of our BD 
simulation. In this context, we note here that up to several hundreds
multi-scattering processes between beads are involved in one MD step of
our BD method as explained in I.

Another interesting observation is that both $r_n(t)$ and 
$r_{\Delta u}(t)$ in Interval 4 is the most closer, among the four 
intervals, to a Poisson distribution which is observed in equilibrium 
under $E=0$.  Namely, although $R_l(t)$ is still 
relatively larger in this interval just after the chain is released at 
$t\simeq t_{\rm max}$ from trapping by an obstacle(s), 
many s-beads have invaded into the extended part(s) and the chain
contracts dominantly by the entropic effect. When $R_l(t)$ comes closer 
to its local minimum value due to trapping by a new obstacle(s), i.e., 
in Interval 1, $r_n(t)$ and $r_{\Delta u}(t)$ at larger $n$ and $\Delta 
u$ become bigger than those of Interval 4, indicating that some parts of 
the chain are already elongated by the field. 

\begin{figure}
\caption[]{\makebox[\textwidth][l]{
$r_n$ by the BD calculation.}
}
\label{fig:BDC_n-ARs}
\end{figure}

\begin{figure}
\caption[]{\makebox[\textwidth][l]{
$r_{\Delta u}$ by the BD calculation.}
}
\label{fig:BDC-ARs}
\end{figure}

Now let us turn to Figs.~\ref{fig:BDC_n-ARs} and ~\ref{fig:BDC-ARs} where 
$r_n$ and $r_{\Delta u}$ averaged over the whole time window of the BD run 
under different fields are shown. Up to the field strength of $E \simeq 
0.008$ neither $r_n$ nor $r_{\Delta u}$ significantly deviates from a
Poisson distribution, while in $r_n$ under $E \nge 0.016$ a tail clearly 
shows up at large $n$. This clearly indicates the existence of a crossover 
between dynamical regimes with and without the elongation-contraction motion 
at around $E\simeq 0.008$ with $M=160$ and $a_{\rm gel}=20$. Even in the 
regime with the elongation-contraction motion where $r_n$ exhibits a 
significant deviation from a Poisson distribution, $\Delta u$ remains 
in an exponential distribution, though its width moderately increases with 
increasing $E$. This result again indicates that, even in the 
{\it deterministic} regime where the time evolution of averaged $R_l(t)$ 
or $v_{\rm G}(t)$ looks apparently deterministic, defect dynamics is still 
purely stochastic.

\section{Extended Bond Fluctuation Method}\label{sec:ebfm}

The stochastic nature of `defect' dynamics found by the BD method in the
present work was just the basic assumption {\it a priori} introduced in
the extended bond fluctuation method (EBFM) when we developed 
it.~\cite{azuma99a} The
conventional BFM (CBFM)~\cite{carm88} is a well-established MC method
which simulate thermodynamic properties of polymers by a coarse-grained
chain model on a lattice. It consists of a set of local movements which
suffice the conditions necessary for the chain to evolve by
self-avoiding random walk. In addition to this local updating process we
introduced non-local movements of `defects' in the EBFM. For this
purpose `defects', or `s-monomers' are defined in almost the same way as
s-beads are defined in the present work. We then let s-monomers hop
non-locally within an extended portion of the chain that each s-monomer
lies. The movements are incorporated in such a way that they fulfill the
detailed valance and the self-avoiding conditions. In the EBFM this
non-local movement of s-monomers and the local updating of the CBFM are
tried alternatively. We note that the number of s-monomers does not
change except at the chain ends by the former process, but it does
change anywhere in the chain by the latter updating.

In Fig.~\ref{fig:histgem} we show $r_n$ and $r_{\Delta u}$ of a chain
with $M=80$ obtained in the EBFM under $\Theta=0$, where we denote the
normalized field as $\Theta\equiv qE_{\rm b}/k_{\rm B}T$ for the MC
analysis. As is the case for the BD model, both $r_n$ and $r_{\Delta u}$
obtained by the EBFM obey a Poisson distribution except for regions 
$n,\ \Delta u \nle 5$. Furthermore, in the EBFM, their widths
coincide with each other at large $n$ and $\Delta u$. In this context,
it is noted that we have adopted equal probability of semi-local
movement of an s-monomer by any distance $k$ among possible $n$ ones. 
The width of the distribution $r_n$ of the EBFM is about a half of that
of the BD method. Correspondingly, `defect' density $\rho\ (\simeq
0.33)$ of the EBFM is about twice larger than that of the BD method. 
Although there exist such quantitative differences between the two
methods, the qualitative agreement in the results demonstrated in 
Figs.~\ref{fig:histdem} and \ref{fig:histgem} supports our EBFM approach
in Ref.~\onlinecite{azuma99a}. 

\begin{figure}
\caption[]{\makebox[\textwidth][l]{
The normalized histograms $r_n$ and 
$r_{\Delta u}$ under $\Theta=0$ obtained by the EBFM. }
}
\label{fig:histgem}
\end{figure}

The elongation-contraction motion under CFGE is also observed by the
EBFM. The normalized histograms  $r_n(t)$ and $r_{\Delta u}(t)$ obtained
in the four intervals introduced in Sec.~\ref{sec:def-dyn} are shown in
Figs.~\ref{fig:hst-r-BFM} and \ref{fig:hst-u-BFM}. Their overall
features qualitatively agree with those obtained by the BD calculation. 
By a  detailed inspection of Fig.~\ref{fig:hst-r-BFM}, however, we see
that $r_n(t)$ in Intervals 2
and 3 by the EBFM still exhibit nearly exponential distributions at 
larger $n$ in contrast to those of the BD results which have an enhanced
tail at $n$ as large as several tens. This difference is considered to
originate from the following process of the EBFM: s-monomers are
created or annihilated, and so length $n$ of extended parts are changed,
only through local updating of monomers originally incorporated by the
CBFM. By such local stochastic processes an extremely long extended part
is hardly created.  

\begin{figure}
\caption[]{\makebox[\textwidth][l]{
The normalized histogram $r_n(t)$ obtained by the EBFM.}
}
\label{fig:hst-r-BFM}
\end{figure}

\begin{figure}
\caption[]{\makebox[\textwidth][l]{
The normalized histogram $r_{\Delta u}(t)$ obtained by the EBFM.}
}
\label{fig:hst-u-BFM}
\end{figure}

The normalized histogram $r_{\Delta u}(t)$ shown in
Fig.~\ref{fig:hst-u-BFM}, on the other hand, exhibit exponential
behavior at $\Delta u \nge 2$ quite similar to the corresponding BD
result shown in Fig.~\ref{fig:h-as-BDC}. This is also the case for the
integrated distributions $r_{\Delta u}$ under various field strengths. 
They are shown in Fig.~\ref{fig:BFM-ARs} which one can compare with the 
corresponding BD result in Fig.~\ref{fig:BDC-ARs}. Quantitatively,
$r_{\Delta u}(t)$ and $r_{\Delta u}$ by the EBFM have significant weight 
at relatively larger $\Delta u$ as compared with those by the BD
calculation. This difference is attributed to the process in the EBFM
that, once a long extended part of length $n$ is created, a tangential
move of the s-monomer by the distance $n$ is accepted by the probability 
proportional to $1/n$. Although there exist such minor differences
between the dynamical processes involved in the EBFM and the BD
method, the results demonstrated above are sufficient for us to
conclude that the EBFM properly takes into account the stochastic dynamics
of slack segments which give rise to the elongation-contraction motion
of a polymer under gel electrophoresis.

\begin{figure}
\caption[]{\makebox[\textwidth][l]{
$r_{\Delta u}$ in the EBFM with various field strength $\Theta$.}
}
\label{fig:BFM-ARs}
\end{figure}

\section{Conclusion}\label{sec:conc}

In equilibrium under a vanishing field it is found, as expected, that
s-beads, or `defects', are distributed randomly. Both distributions 
of extended parts $r_n(t)$ and of their tangential
movements $r_{\Delta u}(t)$ obey a Poisson distribution. In the
relatively strong field region where a chain exhibits the
elongation-contraction motions giving rise to quasi-periodic
behavior in $R_l(t)$ and $v_{\rm G}(t)$, $r_n(t)$ is observed to show
the corresponding time evolution. Especially, immediately before
$R_l(t)$ reaches to a local maximum, chain conformations with extended
parts of more than a few tens of beads, which are entropically
unfavorable, are observed. As for $r_{\Delta u}(t)$, such a large
fluctuation seems not to occur. We therefore conclude that
the s-bead dynamics at a semi-microscopic level is rather stochastic,
though the coarse-grained quantities such as $R_l(t)$ look apparently
deterministic during the elongation-contraction motion, or in time
ranges which we have called {\it deterministic} ones.
We have also shown that these features of $r_n(t)$, $r_{\Delta u}(t)$
and $R_l(t)$ are similar to those observed by the MC analysis based on
the EBFM which we recently proposed. Such coincidences between the two
simulations strongly support our algorithm of EBFM which requires an
order of magnitude less CPU times than MD calculations based on the 
present BD method do.

\begin{acknowledgments}

The author wish to thank H. Takayama for useful discussions and
suggestions. The computation in the present work has been done using the
facilities of the Supercomputer Center, Institute for Solid State Physics, 
University of Tokyo, and those of the Computer Center of University 
of Tokyo.

\end{acknowledgments}

\newpage
\bibliographystyle{apsrev}
\section*{References}

\newpage
\printfigures

\newpage
\begin{center}
\resizebox{!}{7mm}{FIG.~1}\\\vspace*{25mm}
\leavevmode\epsfxsize=80mm
\epsfbox{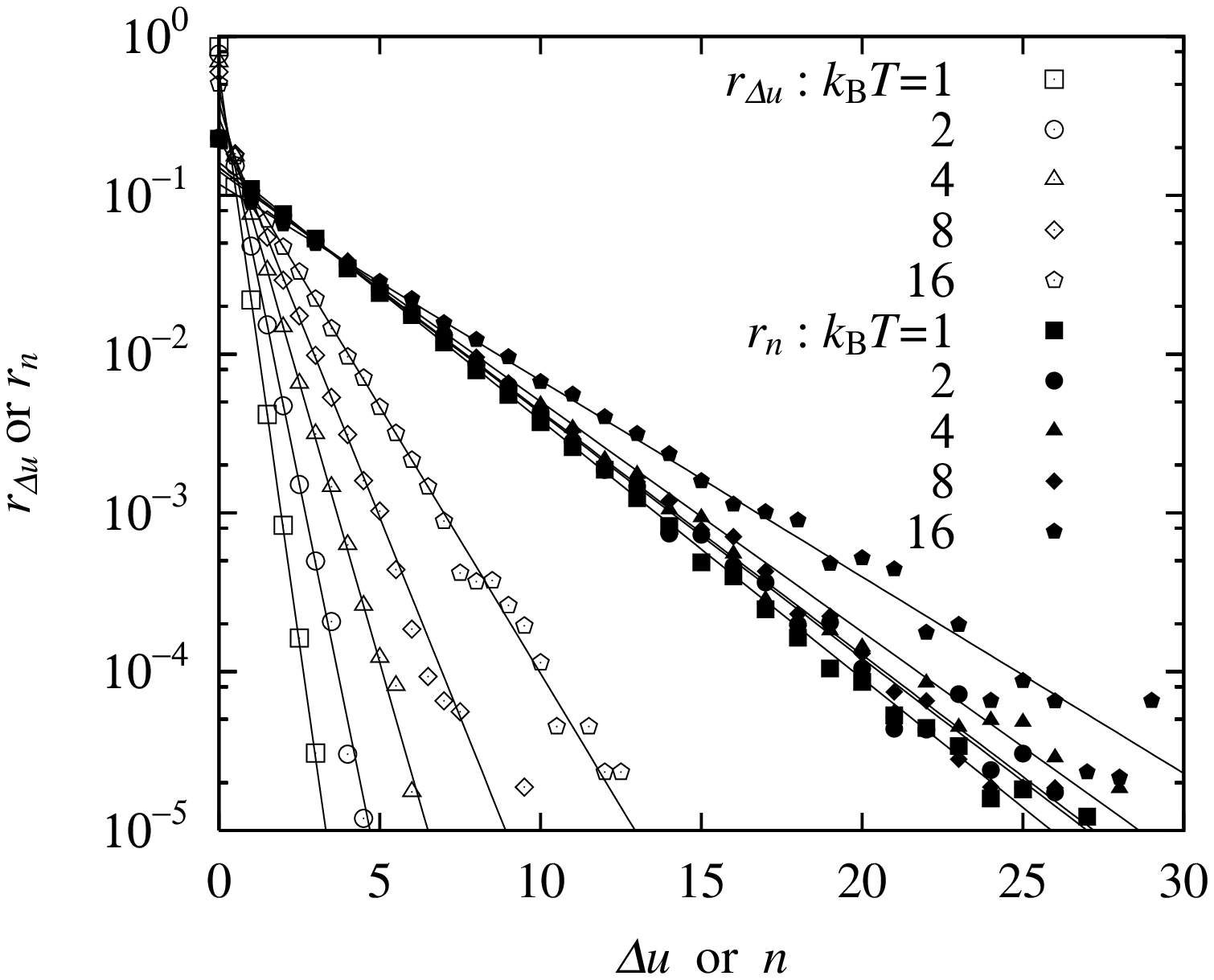}
\end{center}
\clearpage

\newpage
\begin{center}
\resizebox{!}{7mm}{FIG.~2}\\\vspace*{25mm}
\leavevmode\epsfxsize=80mm
\epsfbox{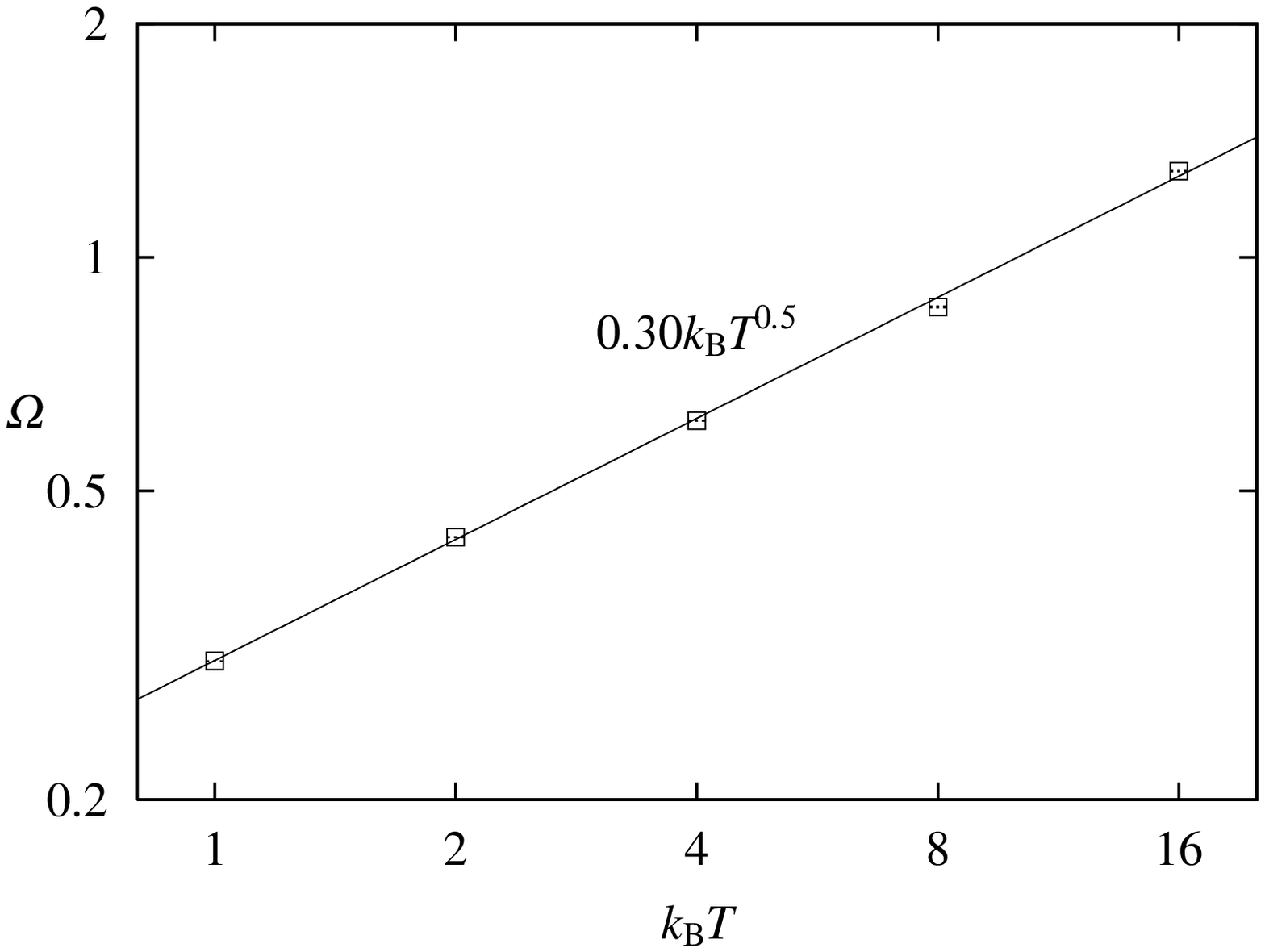}
\end{center}
\clearpage

\newpage
\begin{center}
\resizebox{!}{7mm}{FIG.~3}\\\vspace*{25mm}
\leavevmode\epsfxsize=80mm
\epsfbox{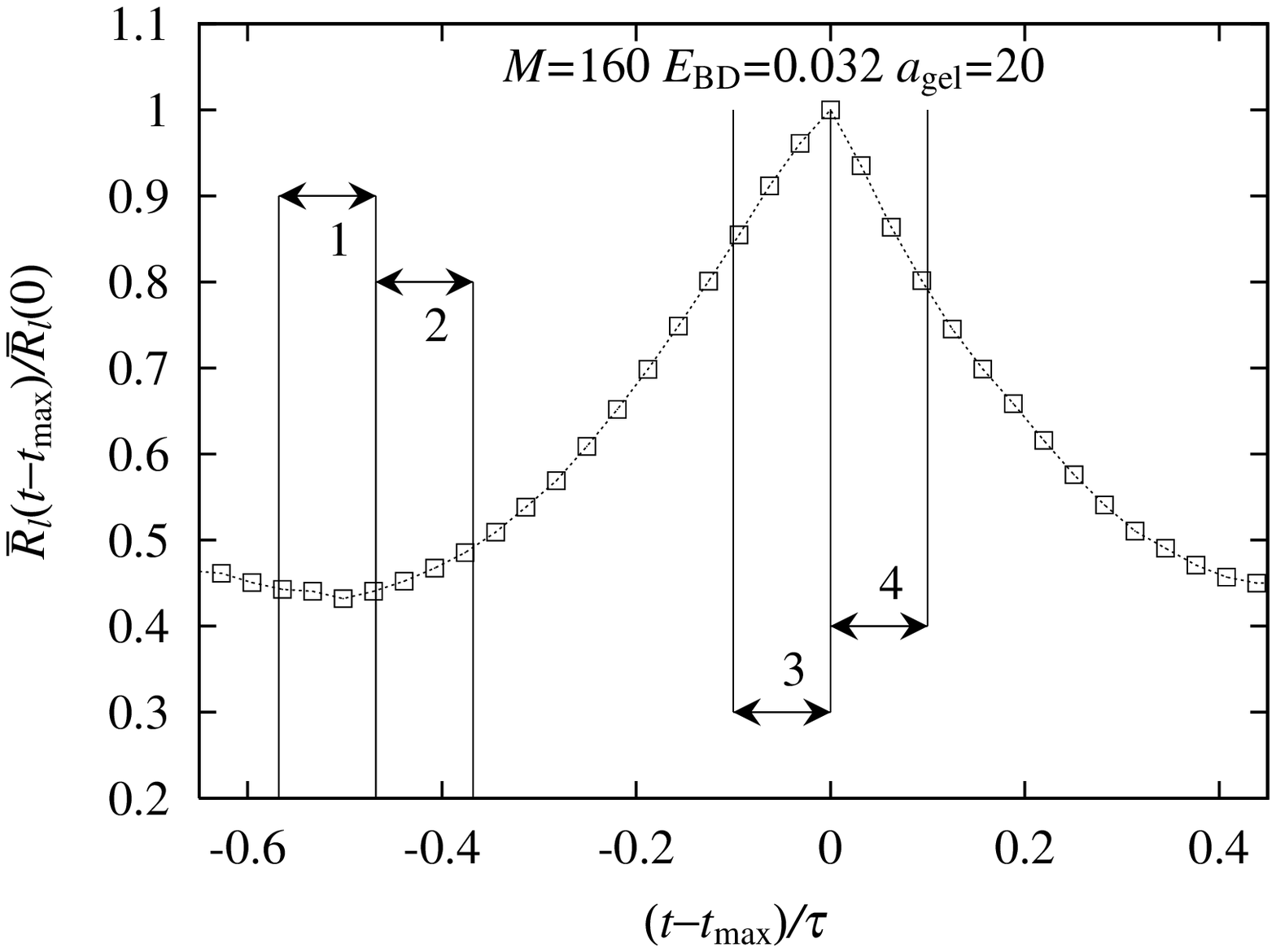}
\end{center}
\clearpage

\newpage
\begin{center}
\resizebox{!}{7mm}{FIG.~4}\\\vspace*{25mm}
\leavevmode\epsfxsize=80mm
\epsfbox{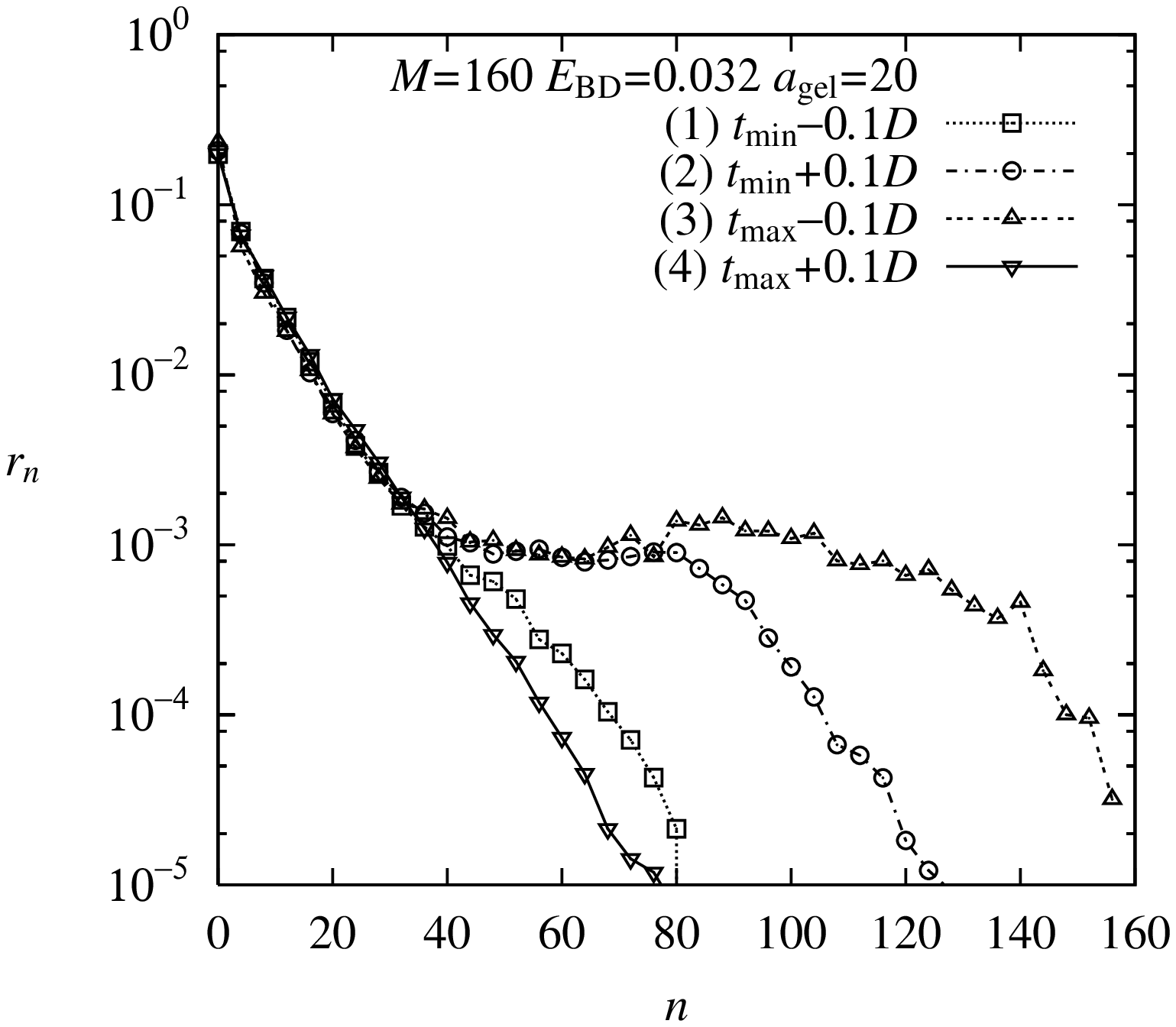}
\end{center}
\clearpage

\newpage
\begin{center}
\resizebox{!}{7mm}{FIG.~5}\\\vspace*{25mm}
\leavevmode\epsfxsize=80mm
\epsfbox{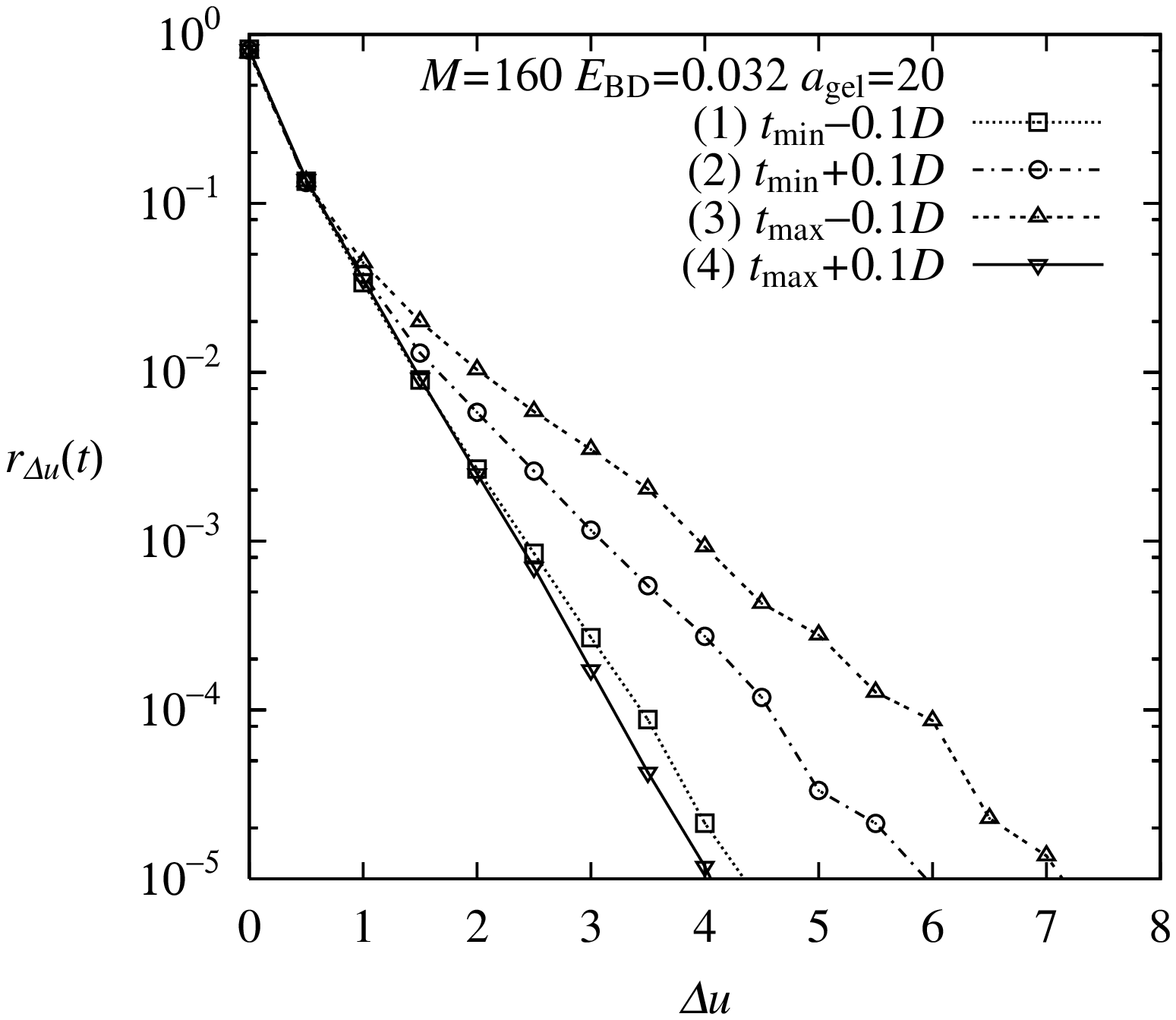}
\end{center}
\clearpage

\newpage
\begin{center}
\resizebox{!}{7mm}{FIG.~6}\\\vspace*{25mm}
\leavevmode\epsfxsize=80mm
\epsfbox{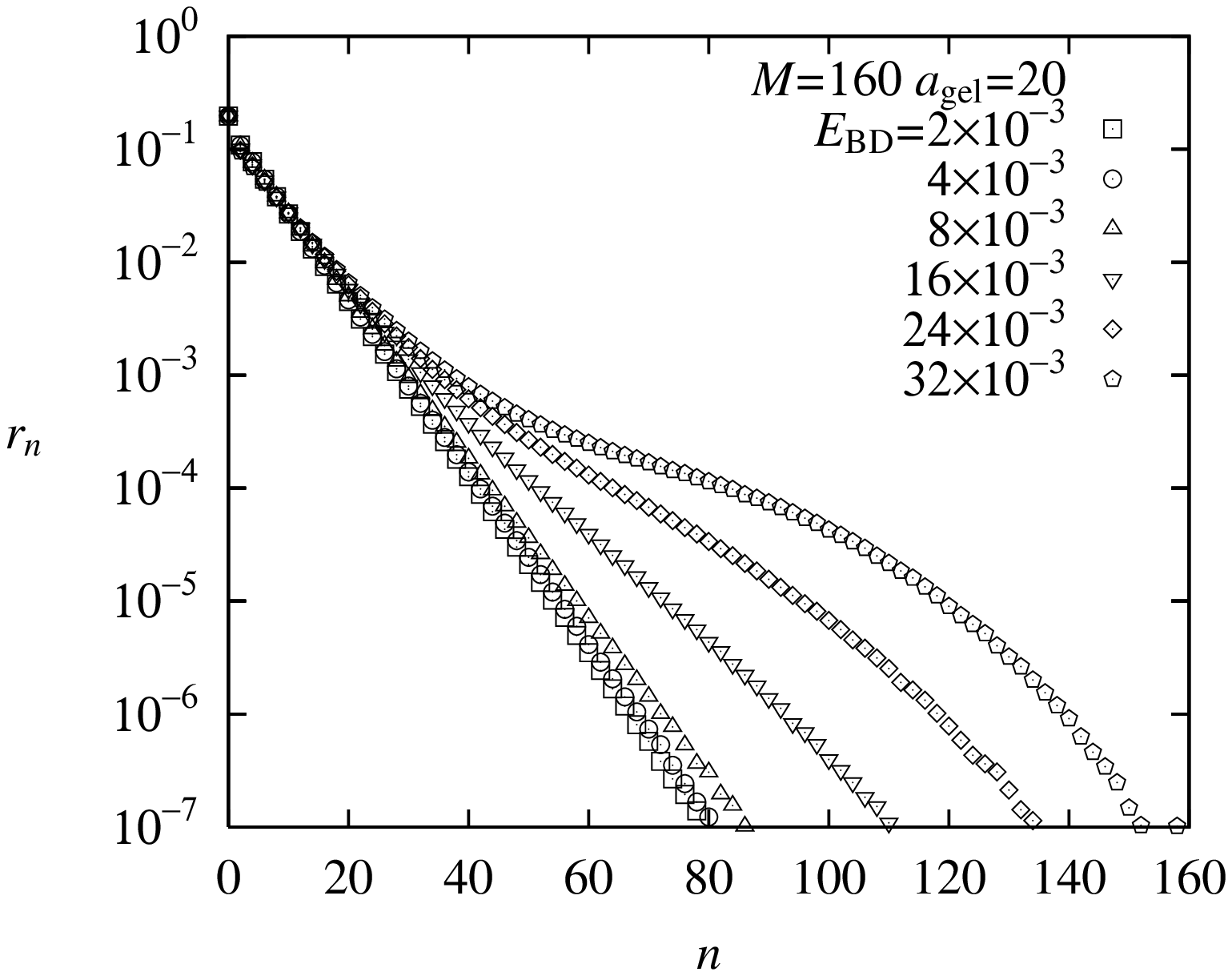}
\end{center}
\clearpage

\newpage
\begin{center}
\resizebox{!}{7mm}{FIG.~7}\\\vspace*{25mm}
\leavevmode\epsfxsize=80mm
\epsfbox{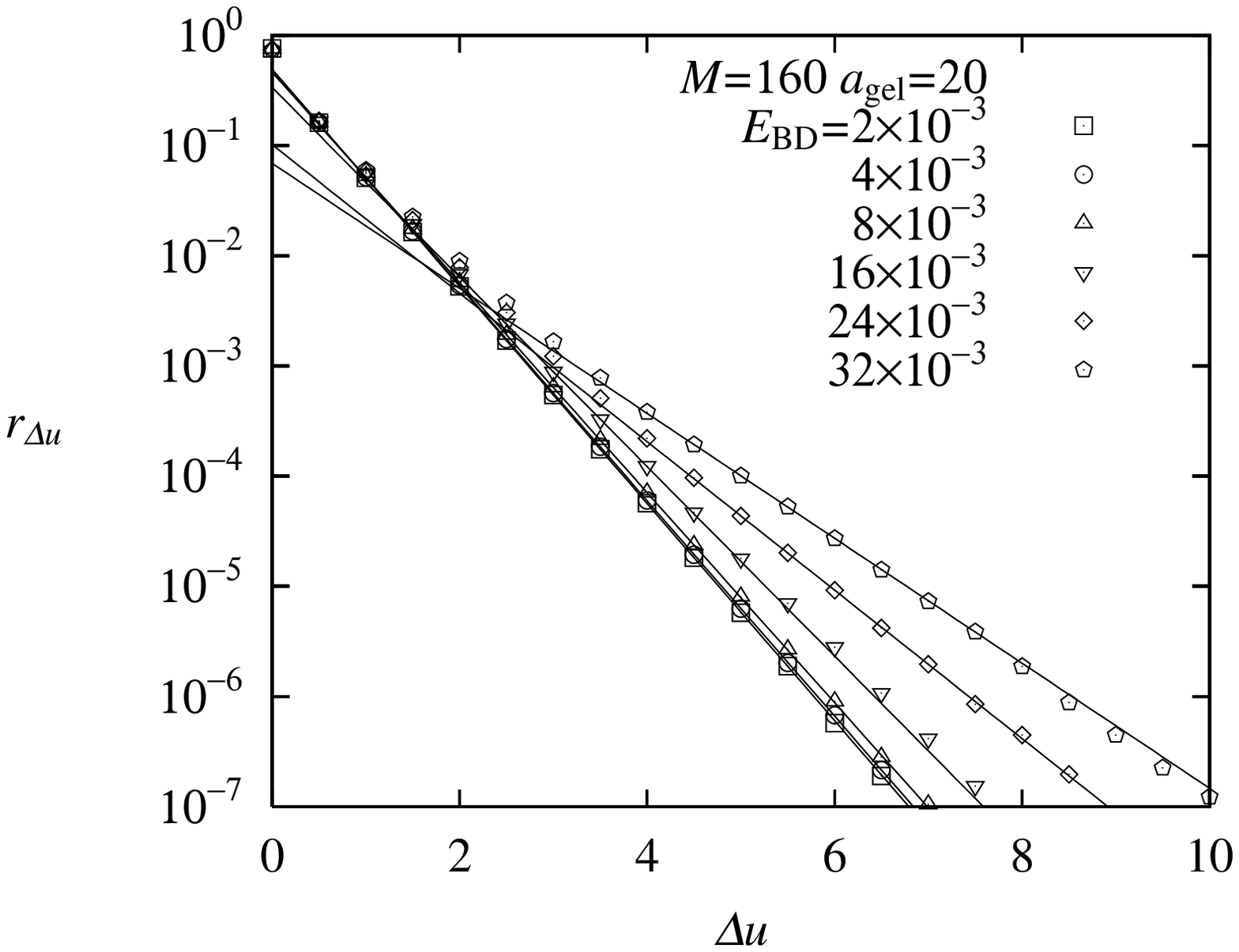}
\end{center}
\clearpage

\newpage
\begin{center}
\resizebox{!}{7mm}{FIG.~8}\\\vspace*{25mm}
\leavevmode\epsfxsize=80mm
\epsfbox{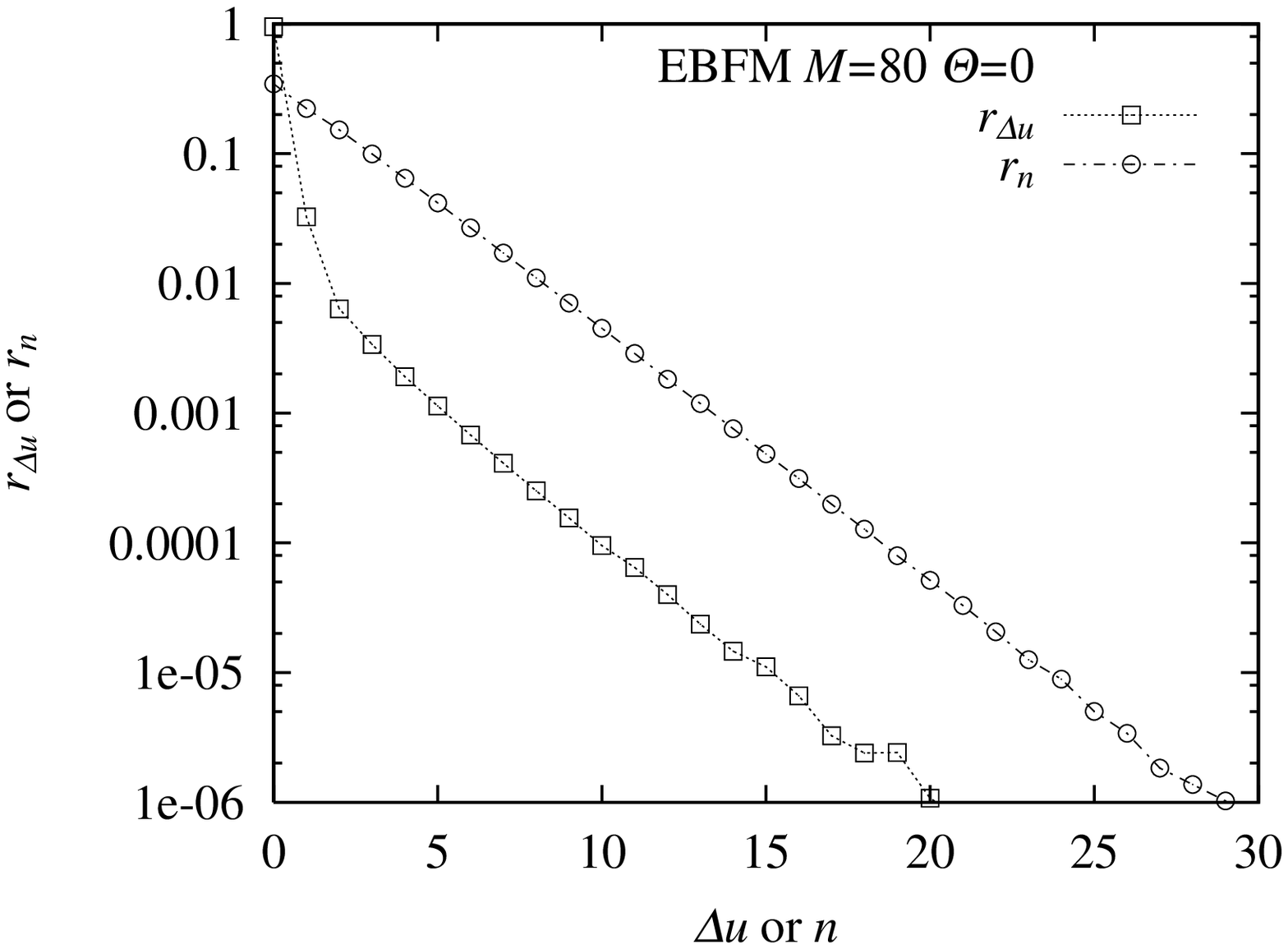}
\end{center}
\clearpage

\newpage
\begin{center}
\resizebox{!}{7mm}{FIG.~9}\\\vspace*{25mm}
\leavevmode\epsfxsize=80mm
\epsfbox{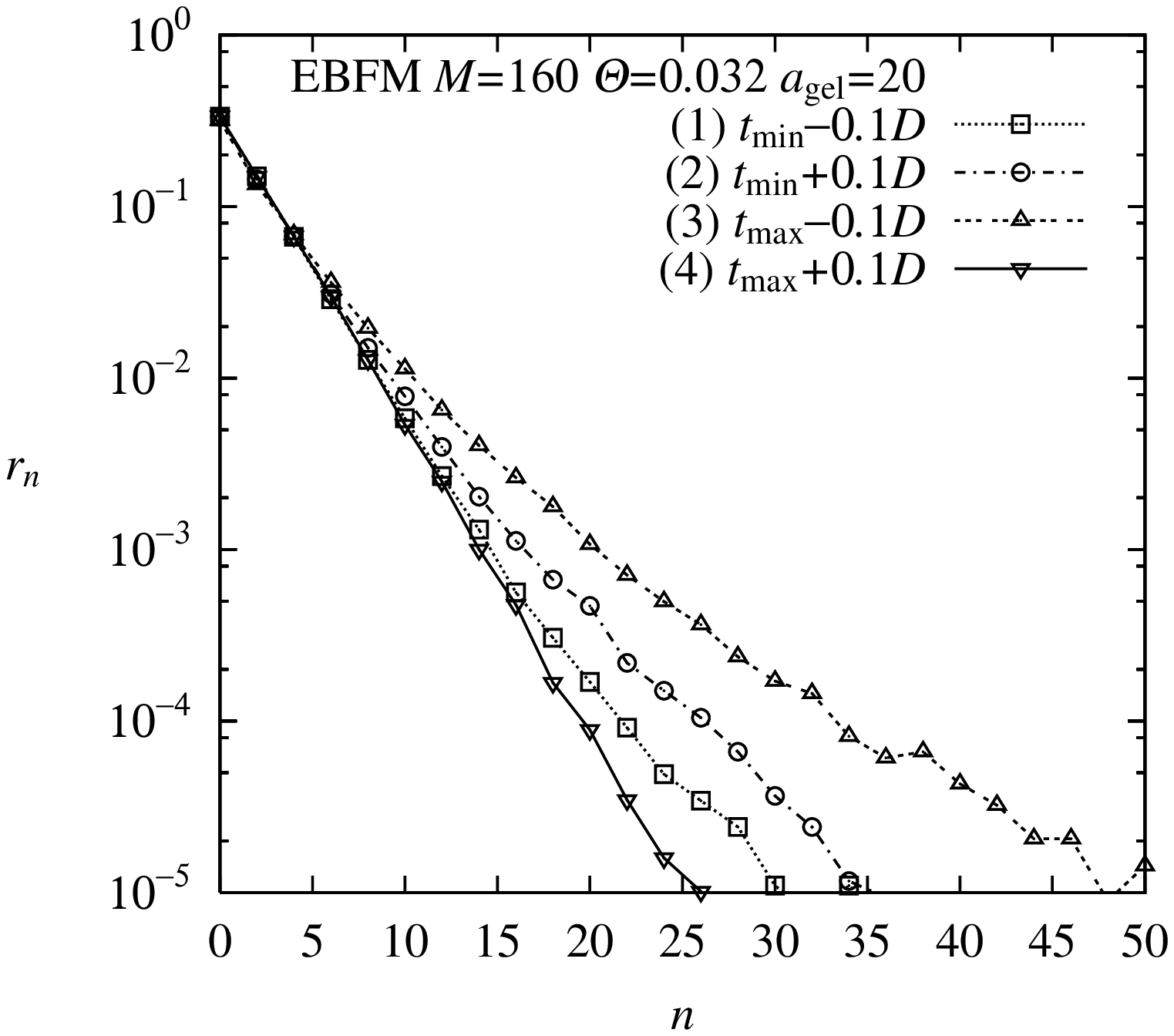}
\end{center}
\clearpage

\newpage
\begin{center}
\resizebox{!}{7mm}{FIG.~10}\\\vspace*{25mm}
\leavevmode\epsfxsize=80mm
\epsfbox{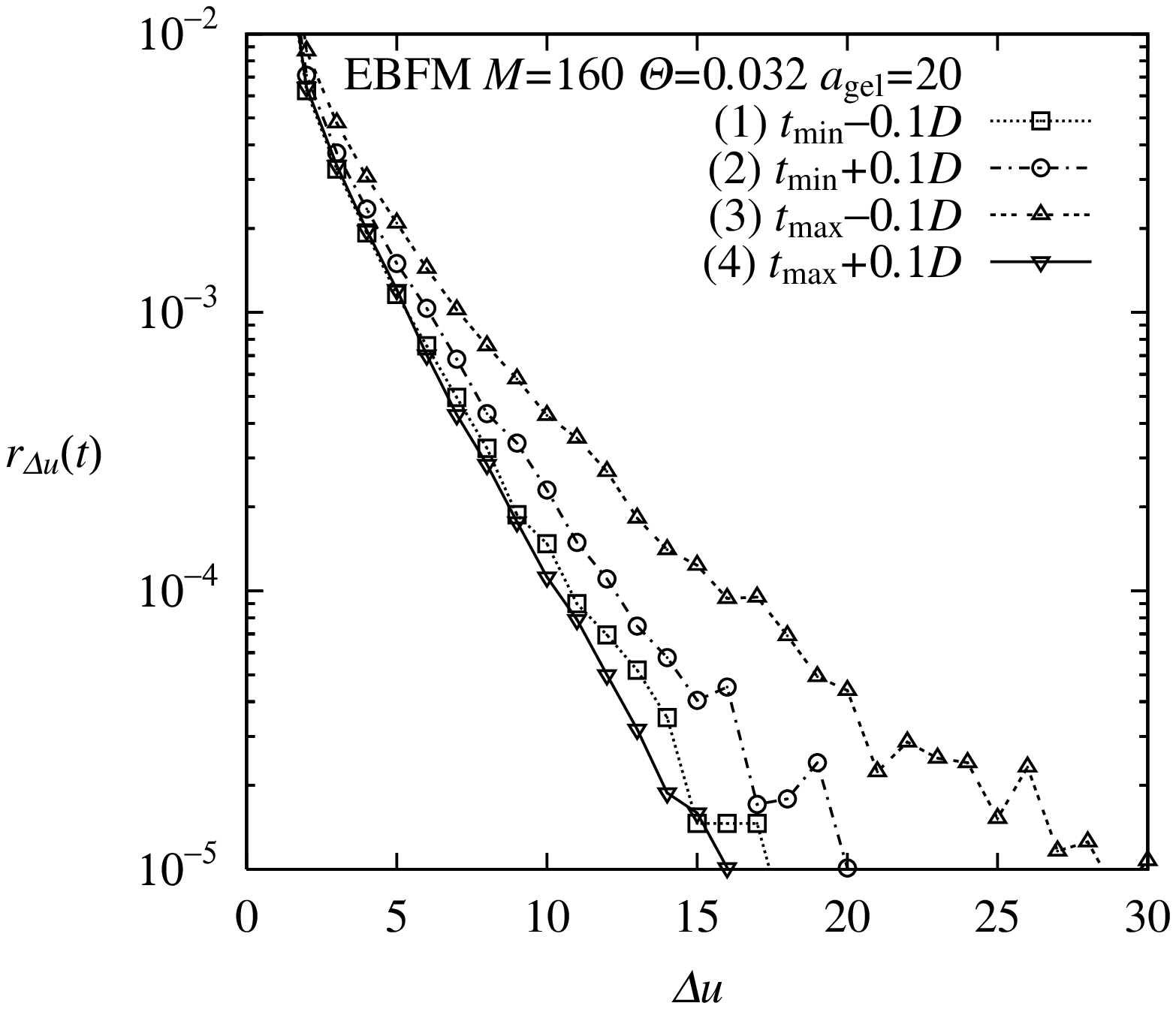}
\end{center}
\clearpage

\newpage
\begin{center}
\resizebox{!}{7mm}{FIG.~11}\\\vspace*{25mm}
\leavevmode\epsfxsize=80mm
\epsfbox{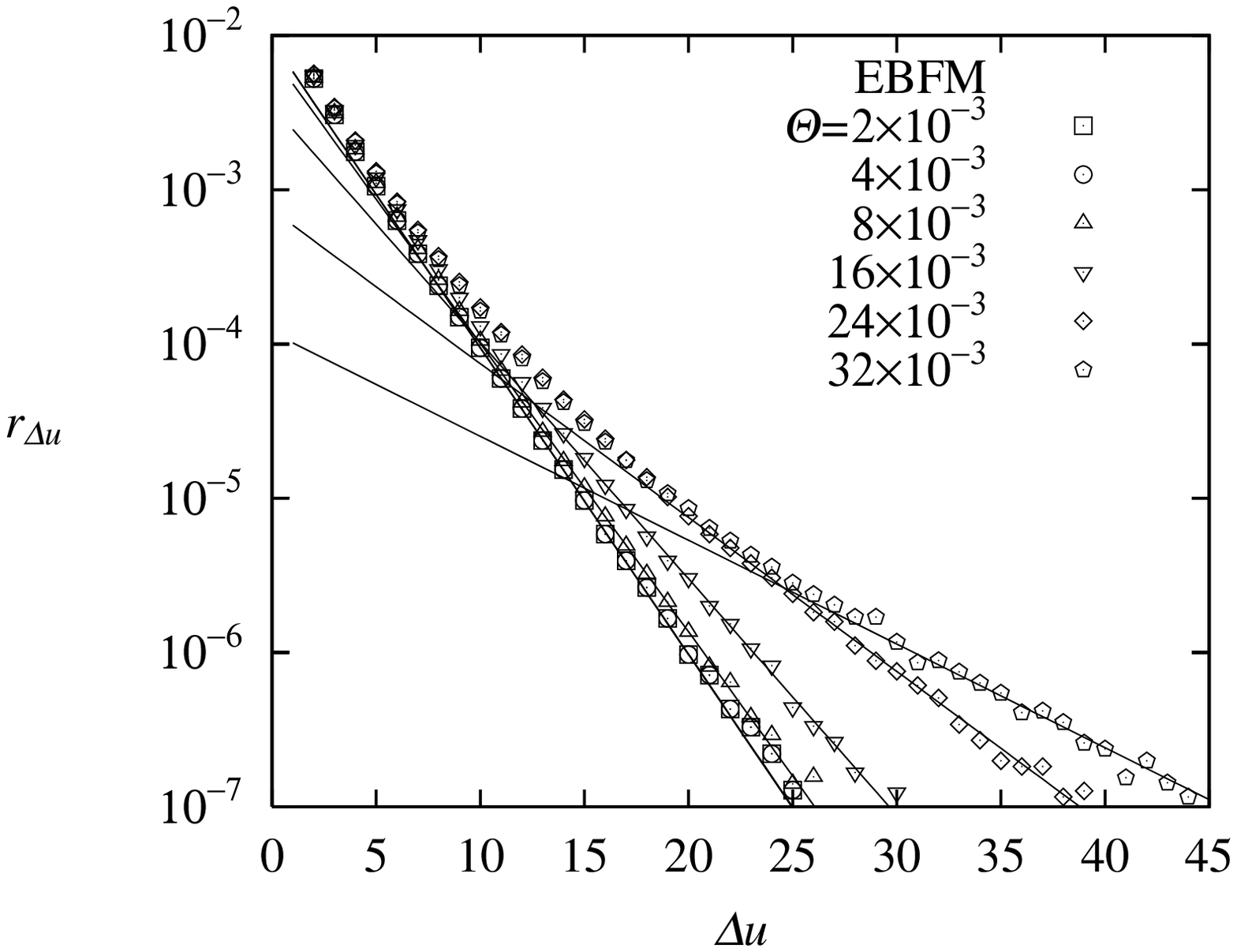}
\end{center}
\clearpage

\end{document}